# Thermal Recovery Mechanisms of UO$_2$ Lattices

Seçkin D. Günay

Yıldız Technical University, Department of Physics, Faculty of Science, Esenler, 34210, İstanbul, Turkey

**Abstract**

Thermal recovery of irradiated uranium dioxide lattice have been investigated by experimental groups and defects related to each recovery stages are estimated based on the indirect observations like activation energies. In this study, thermal recovery mechanisms of both neutron and alpha-particle irradiated uranium dioxide lattice is investigated by atomistic simulation methods. Isochronal and isothermal annealing have been applied to the irradiated samples. Following the same procedures of the experimental groups, activation energies are calculated with molecular dynamics simulations and compared with empirical data which show a good agreement. Migration energy barrier and the recovery energies of the obstruction type defects are calculated with molecular dynamics simulation and density functional theory. Recovery stages of the irradiated crystals are discussed however unlike the predictions of the most experimental groups, it is shown that neutron and alpha-particle irradiated crystal structures do not thermally recover for the same reason. All three recovery stages of neutron irradiated uranium dioxide crystal is attributed to oxygen interstitials while the three recovery stages of the alpha-particle irradiated crystals is considered to the oxygen – uranium, uranium and uranium interstitial defects respectively. Finally, the reason and the main driving force of thermal recovery of irradiated crystals is the obstruction type defects which give a lead to all other types of defects recover with themselves in each lattice recovery stages.

## 1. Introduction

In the previous study [1], swelling mechanisms of the uranium dioxide crystal with ingrowth of irradiation induced defects is studied by molecular dynamics (MD) simulation method and compared with experimental results [2,3,4]. Neutron and alpha-particle irradiation induced defects are investigated and it was shown that main reason for the lattice swelling is the obstruction type defects. They are oxygen ions for neutron irradiated and uranium ions for alpha-particle irradiated crystals.

In this study, damaged uranium dioxide crystal of the previous study is used and their recovery mechanisms is examined by both molecular dynamics and density functional methods. Time and temperature has great influence on the damaged crystals, for this reason isochronal and isothermal recovery behavior of the lattice parameter change is investigated and compared with the experimental results. Analogous to the experiments three recovery stages are observed and activation energies are obtained for each stage. Experimental groups suggested defects for each recovery stages and large degree of consensus considers recovery stages are oxygen interstitials, uranium vacancies and helium ions trapped in vacancies, respectively [5-8]. In this work, recovery stages are observed and for each stage main driving force for the recovery of the lattices are the obstruction type defects. In alpha-particle irradiated crystals, as the obstruction type uranium interstitial defects are combining with the uranium vacancies, highly distorted oxygen ions also migrate to their normal positions as a result both oxygens and uraniums are recovered in the first stage. In the second and third stages obstruction type interstitial uranium ions are recovered. Here it could be said that, oxygen migration in the first and helium migration in the third stage is a side effect of recovered uraniums. Contrary to the ideas put forward by experimentalists, in the neutron irradiated uranium dioxide crystals, oxygen interstitials are recovered in all stages. Finally, as the recovery take place, migration energy, lattice and energy change per obstruction type defect are calculated with two different computational methods: classical molecular dynamics and first-principle methods. Results are compared with each other.

## 2. Computational Methods

### 2.1 Density Functional Theory Calculations

Density functional theory formalism was used to calculate the ground state energies, forces and structures of defected (Uranium Frenkel pair) and defect free $UO_2$ crystals. These calculations are performed with projector augmented-wave (PAW) method which is implemented in the VASP (Vienna Ab initio simulation package) package. $UO_2$ system possesses the localized $5f$ electrons. Localized (strongly correlated) $f$ electrons should not be handled with the same way of delocalized electrons where delocalized ones are exposed to an average Coulomb potential. However there exists strong Coulomb repulsion between localized $f$ electrons and for this reason Hubbard term $U$ is included in the calculations. DFT+$U$ method is used to describe the onsite interactions (between electrons localized on the same atomic center) instead of traditional DFT method.

Generealized gradient approximation (GGA) is choosen for the exchange-correlation approximation. Hubbard parameters are $U$=4.5eV and $J$=0.51eV as it was used in previous studies [9-11]. Cutoff energy of plane wave is 500 eV, 2×2×2 Monkhorst-Pack k point mesh is used for the sampling of the irreducible part of the Brillouin zone. Supercell contains 8 unit cells (2×2×2) with 96 atoms ($U_{32}O_{64}$).

$UO_2$ unit cell has $CaF_2$ type structure with Fm-3m space group. In order to simulate a uranium Frenkel pair defect, uranium ion is removed from its normal site to an interstitial position which is called octahedral site. This point is surrounded by six uranium ions where the interstitial is equidistant from the other six uranium ions which is also called here the obstruction type defect.

## 2.2 Molecular Dynamics Simulations

Newton's equations of motion are solved iteratively for each particle in the supercell. Two types of potential models are used in this study. First one is the pair potential which is proposed by Yakub et al. [12] and the second one is the pair potential model with the many-body EAM (Embedded atom model) description which is proposed by Cooper et al. [13].

The interaction for the first potential between particles i and j is,

$$\phi_{ij}(r_{ij}) = \frac{Z_i Z_j}{r_{ij}} + A_{ij} e^{-r_{ij}/\rho_{ij}} - \frac{C_{ij}}{r_{ij}^6} + D_{ij}\left(e^{-2\beta_{ij}(r_{ij}-r_{ij}^0)} - 2e^{-2\beta_{ij}(r_{ij}-r_{ij}^0)}\right) \quad (1)$$

Here first term is the long-range Coulomb interaction and Z is the charge of the ion. Electrostatic part is calculated by Ewald method and the cutoff distance is taken as 10Å. The remaining terms are short-range interactions: Second and third terms are Buckingham and fourth term is the Morse-type interactions. Parameters are developed by Yakub et al. [12] and clarified by Devanathan et al. [14]. Beeman algorithm is employed for the integration of Newton's equations. Maxwell-Boltzmann distribution is used to assign velocities for each ion at the desired temperature, by this way initial kinetic energy is determined. Initial configuration results an excess potential energy as a result there is too much increase in both kinetic and potential energy. In order to avoid such a problem, system is equilibrated for 30 ps with the time step is 1 fs. However, as the physical properties calculated, this part is not included in the calculations. Total simulation time is 100ps, in the latter part of this article total simulation time is increased to 500ps and finally to 10000 ps. Temperatures of all systems are increased from 300K to 3200K with 50K interval.

NPT ensemble is applied to the system by the program MOLDY [15]; P is the Parinello-Rahman compatible constant pressure method where system is constrained by uniform-dilation, T is the constant temperature ensemble which means system is coupled with a heat bath using Nose-Hoover method.

The second interaction potential [13] has two parts, which are pairwise and many-body parts. Total simulation time is 100ps and the cut-off distance is 11Å [16]. Details of the simulation parameters are given in the paper of Cooper et al. [13].

8×8×8 supercell box is used which contains 2048 Uranium and 4096 Oxygen ions. Defected samples are prepared in the same way as explained in detail in our previous study [1]. A sample figure of supercell with 60 uranium IFP is displayed in Figure 1.

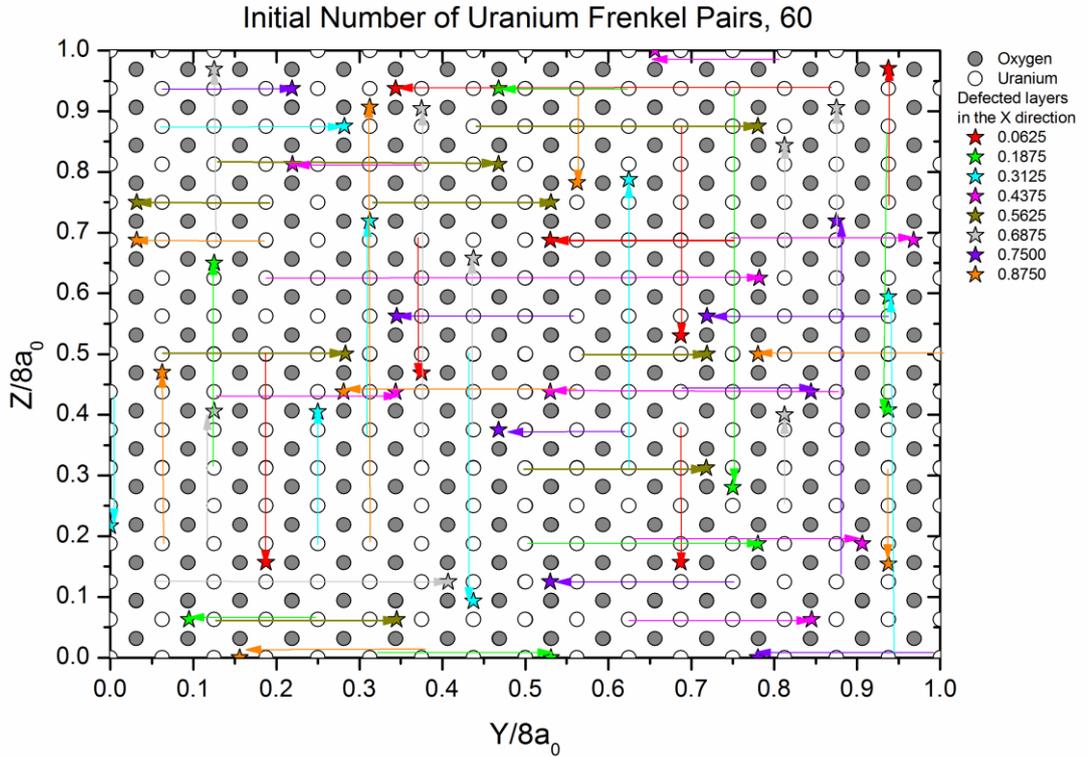

Figure 1. A sample figure of supercell with 60 uranium IFP

Average number of defects are calculated with the following expression in the range Δr around a central ion.

$$\overline{M_\alpha} = \frac{\sum_{i=1}^{n_t}\sum_{j=1}^{N_\alpha} K_{ij}}{n_{\alpha\beta} n_t} \qquad (2)$$

All the variables of the equation are explained before [1]. Δr interval is chosen from the radial distribution function depending on what type defect one is looking for; obstruction/distortion and uranium/oxygen defects. For example obstruction type uranium defects are calculated from Figure 2 for $\Delta r \cong 3.35 - 0.0$ and calculated as $\overline{M_U^{obs}} \cong 44.9$.

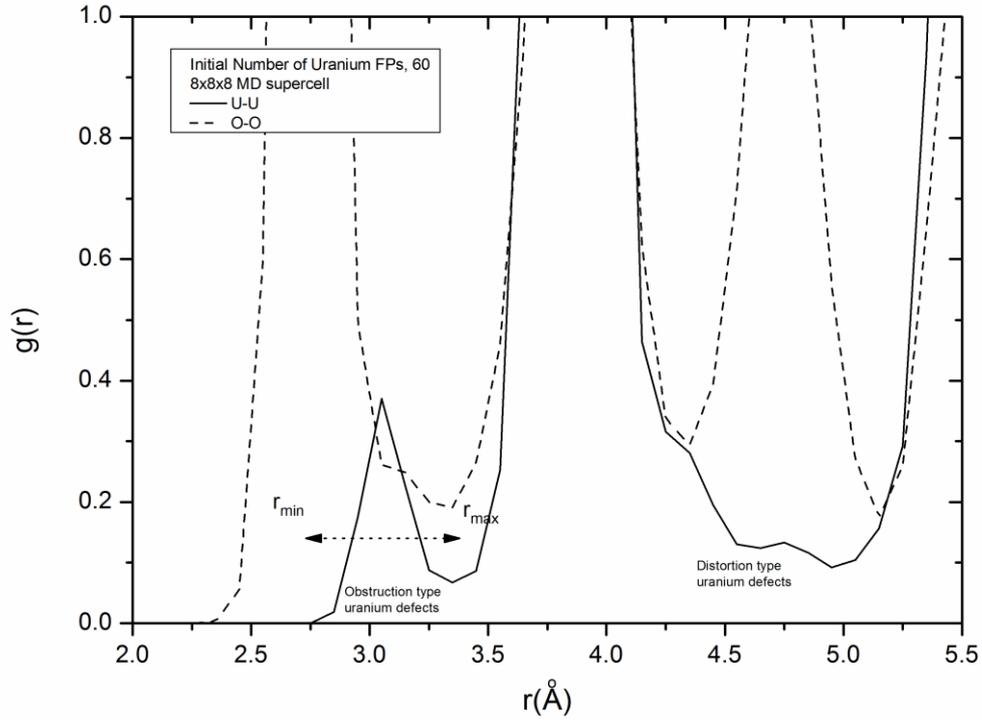

Figure 2. Radial distribution function after equilibrium; sample with 60 uranium IFP.

Unlike distortion type defects, obstruction type defects have more clear boundaries in the radial distribution function (Figure 2). These interstitial defects have a constant number of ions surrounding them. Here uranium type obstruction defect is at octahedral position surrounded by six uranium ions. Oxygen type obstruction defect is surrounded with four uranium and four oxygens. These defects have their names because they obstruct the <110> channel. In addition to this, distortion type have its name because they are distorted into the <110> channel.

## 3. Results and Discussions

In the previous study [1] irradiation of $UO_2$ with neutrons results oxygen type defects and irradiation with alpha particles results both uranium and oxygen type defects. As the defects are created a clear linear dependence of relative lattice expansion versus number of obstruction type defects is observed. For the neutron irradiation these obstruction type defects are oxygen interstitials and for the alpha-particle irradiation they are uranium interstitials.

## 3.1 Isochronal annealing

### 3.1.1 Supercells with Oxygen Frenkel Pair Defects

$UO_2$ single crystal is irradiated with neutrons by Nakea et al. [2, 3] and lattice parameter change is observed (Figure 3a inset). It was shown that [1] as the number of initial oxygen Frenkel pair (IFP) defects is increased lattice parameter increases to a saturation value and there is a resemblance between the experiment and simulation data. In this study, MD simulation re-performed for 8×8×8 supercell of $UO_2$ for the two interaction potentials at 300K (Figure 3a). Nakea et al. observed two stages up to the highest data point and these stages can also be observed in the simulation results. Here the highest data point should be around 80 IFP in Figure 3a.

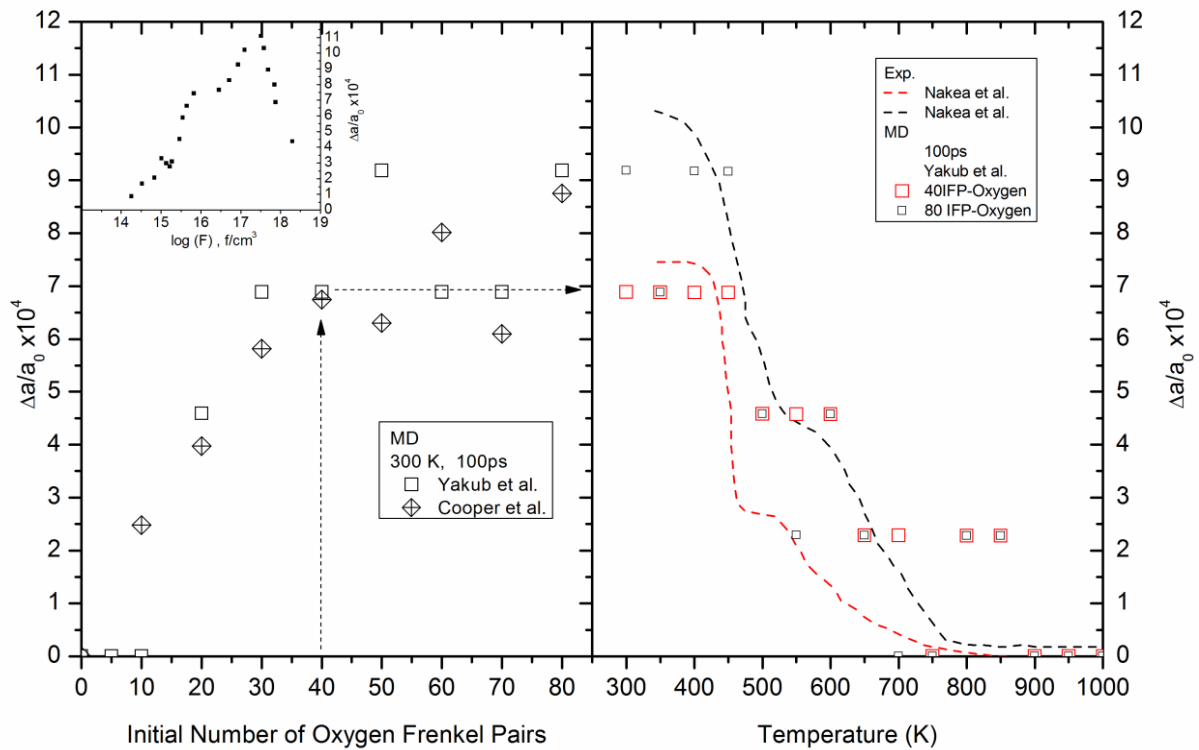

Figure 3. MD simulation results are compared with experimental data of Nakea et al. [5,8] a) Relative lattice increase of $UO_2$ with increasing number of oxygen IFPs. b) Thermal recovery of $UO_2$ lattice for 40 and 80 oxygen IFPs.

Thermal recovery experiment is performed by Nakea et al. [5] for the neutron irradiated $UO_2$ sample (Figure 3b). He suggested the first recovery stage is attributed to the migration of oxygen vacancies and the second stage is due to uranium vacancies. Some other studies [17, 18] commented that first stage should be from the migration of oxygen interstitials. Similar stages are observed with the MD simulation in both samples (IFP 40 and IFP 80) and total recovery temperature is about 900K which is in agreement with the experiment. It could be concluded that annihilation of oxygen Frenkel pairs are the reason for the recovery stages in neutron irradiated $UO_2$ crystal. These oxygens are obstruction type defects, surrounded with four oxygens and four uranium ions. One last note should be added about the precision; in this part very small changes in the lattice parameter is observed in the MD simulation. Clearer peak point with less scattered data could be observed with bigger supercells and performing MD simulations for more data points in Figure 3. Besides, equidistant jumps of the lattice parameter data of Yakub et al. potential is from the precision limits of the MD program [15]. These could not be observed with the Cooper et al. potential because other MD program [16] have higher precision. In the following part such a high precision would not be necessary because lattice expansion with alpha-particle irradiation is almost ten times larger than the neutron irradiation.

### 3.1.2 Supercells with Uranium Frenkel Pair Defects

$UO_2$ single crystal is irradiated with alpha particles by Weber [4] and an exponential growth of lattice parameter change is observed (Figure 4a inset). In previous study [1], it was put forward that there is a correlation between the relative change in the lattice parameter by the alpha irradiation dose of $UO_2$ from the experimental study and the relative change in the lattice parameter by the increase in the numbers of uranium IFP defects from the MD simulation (Figure 4a). Moreover it was concluded that obstruction type uranium FP defects (interstitial uranium ions located at octahedral sites) are the main reason of lattice swelling and they are linearly dependent which means as the number of obstruction type uranium FP defects increases lattice parameter changes linearly to a saturation point.

Weber also investigated the recovery of alpha particle irradiated single crystal $UO_2$ by increasing the temperature. Three recovery stages are observed in his experimental study at 573K, 848K and 1198K (Figure 4b and Figure 5). He also found that total recovery is around 1323K [6 Weber 1983]. Above 1323 K Weber considers total recovery take place where the last data point is $\Delta a/a_o = 0.06\%$.

Analogously, this work will focus on the recovery behavior of both the lattice parameter and defects as the temperature is increased by using isochronal and isothermal annealing.

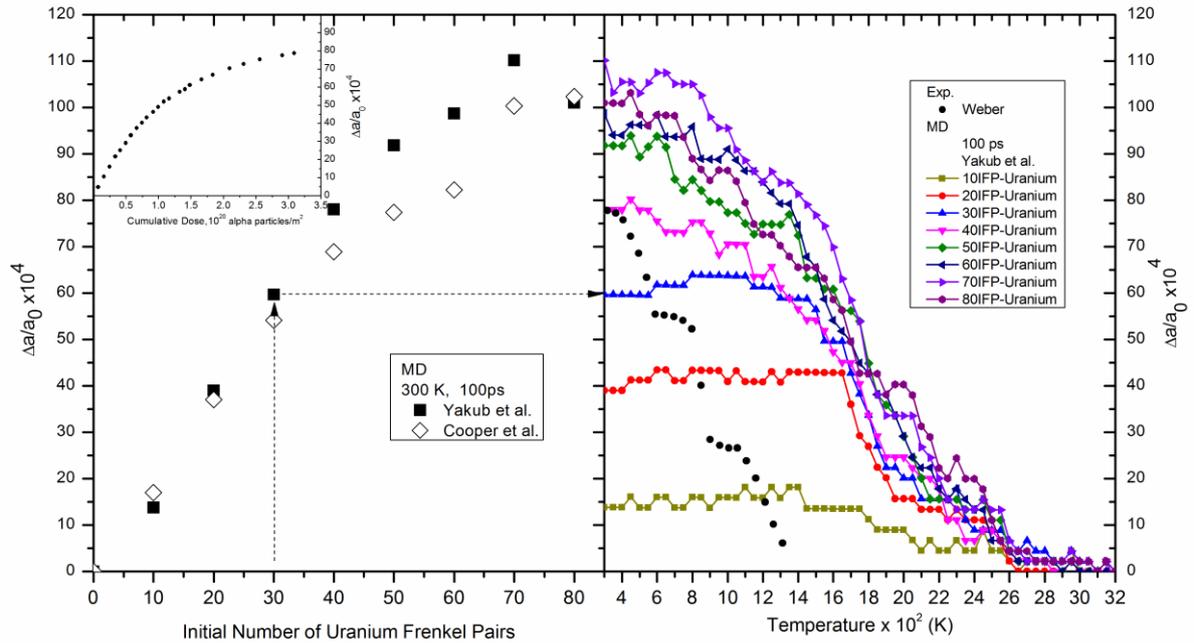

Figure 4. MD simulation results are compared with experimental data of Weber [4, 6] a) Relative lattice increase of $UO_2$ with increasing number of uranium IFPs. b) Thermal recovery of $UO_2$ lattice for each uranium IFPs.

In Figure 4b lattice parameter recovery of each sample (supercells with different uranium IFPs) is given as the temperature increase. Each sample demonstrates the isochronal recovery. Isochronal recovery data points are obtained from 100ps simulation. Here recovery steps are not clear due to too much scattering in the data. Total recovery occurs around 2400K-2600K ($\Delta a/a_o = <0.06\%$) which is simulation overestimates the experimental value about 1000K.

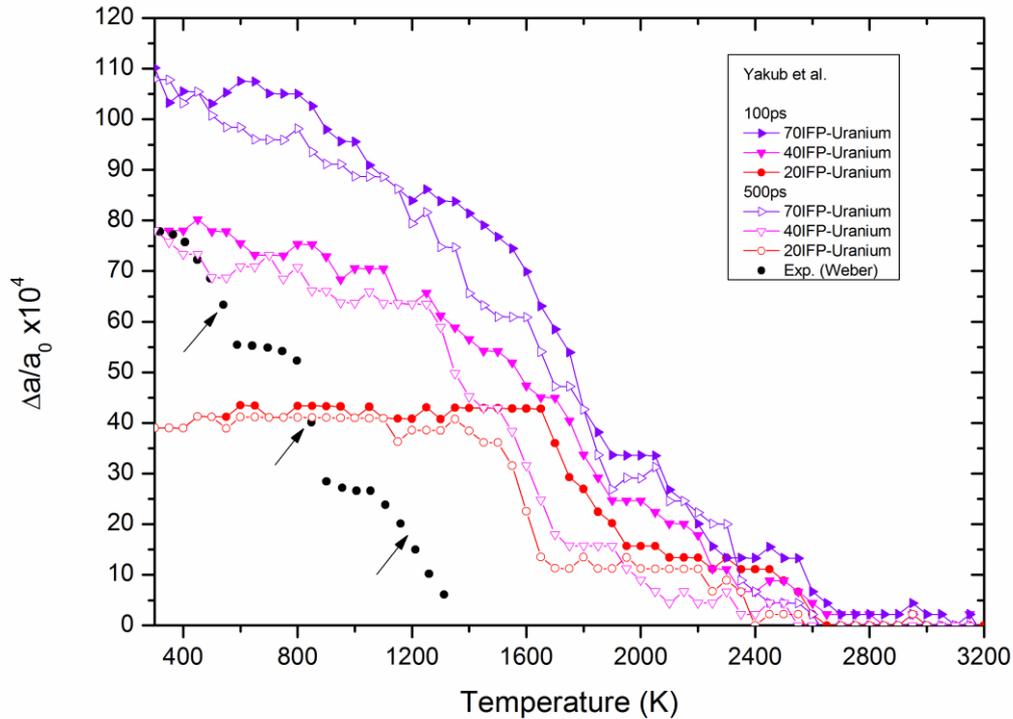

Figure 5. Comparison of MD simulation results of thermal recovery of UO$_2$ lattice for 100ps and 500ps. Arrows indicate the thermal recovery stages of the experiment.

However time is an important variable for recovery so to improve results, simulation time is increased to 500 ps and causes less scattered data with more clear recovery stages. Moreover increasing the total simulation time reduces the overall data points towards experimental value (Figure 5). Total recovery temperature of all samples diminished about 200K - 400K and it is between 2000K - 2400K for 500ps simulation time. Nevertheless it is early to reach a conclusion like, if one increases the time sufficiently large than one could obtain the experimental results. Main reason to avoid such a conclusion could be speculated as insufficient interaction potential or other reasons specific to molecular dynamics methods. Moreover time has a great impact on the recovery mechanism and as it passes concentration of defects in the simulation box could be much lower than the experimental value. Relevant recovery time for simulation could be verified by the stage levels in Figure6.

In figure 6, locations of the stages from the simulation results are indicated for all samples, even they are not clear. For 20 IFP sample two stages and for 40 and 70 IFP samples three stages exist as shown in Figure 6. Temperatures of all stages are overestimated however stage levels, relative lattice expansion (y axis data), are quite close to experimental values

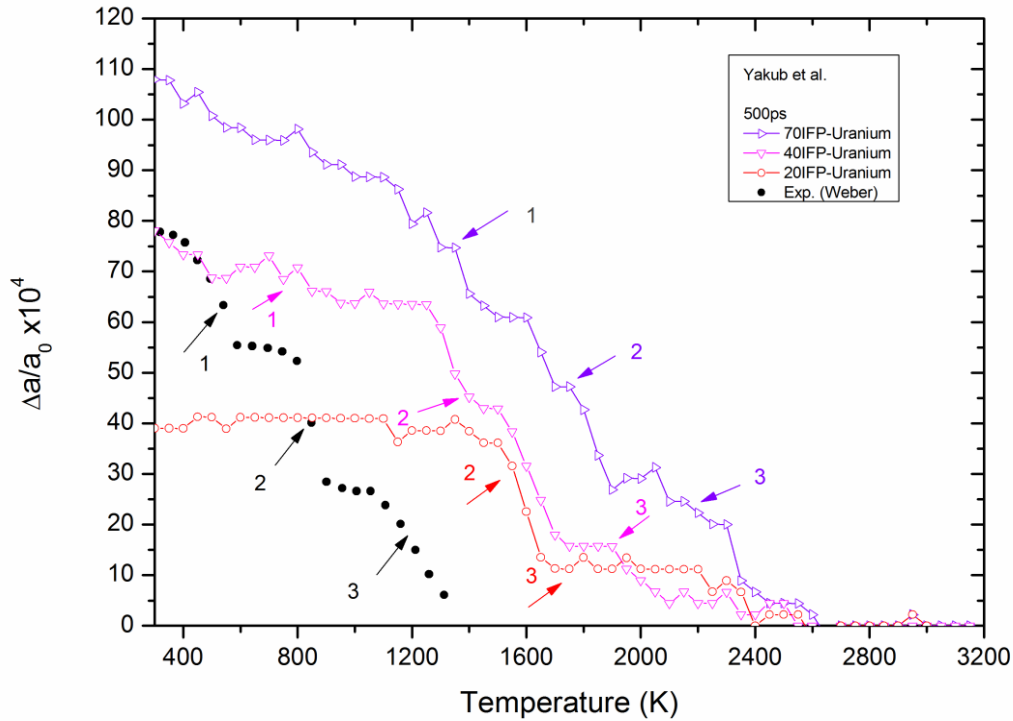

Figure 6. Recovery stages of MD simulation and experimental results.

Weber mentioned the stage 2 recovery is much sharper than the others and with this observation he concludes that in the first and third stage there is a distribution in the activation energies but for the second there is only one activation energy value. In the MD simulation also, stage 2 displays a sharp decrease.

In Figure 6, coordinates of recovery stages are hardly determined so the derivative of the smoothed data of IFP70 (Figure 7a) is plotted in Figure 7. From this new data recovery stages are at points 1300K, 1750K and 2300K. Recovery stage points will also be used in the next section to determine activation energy values from the isothermal data.

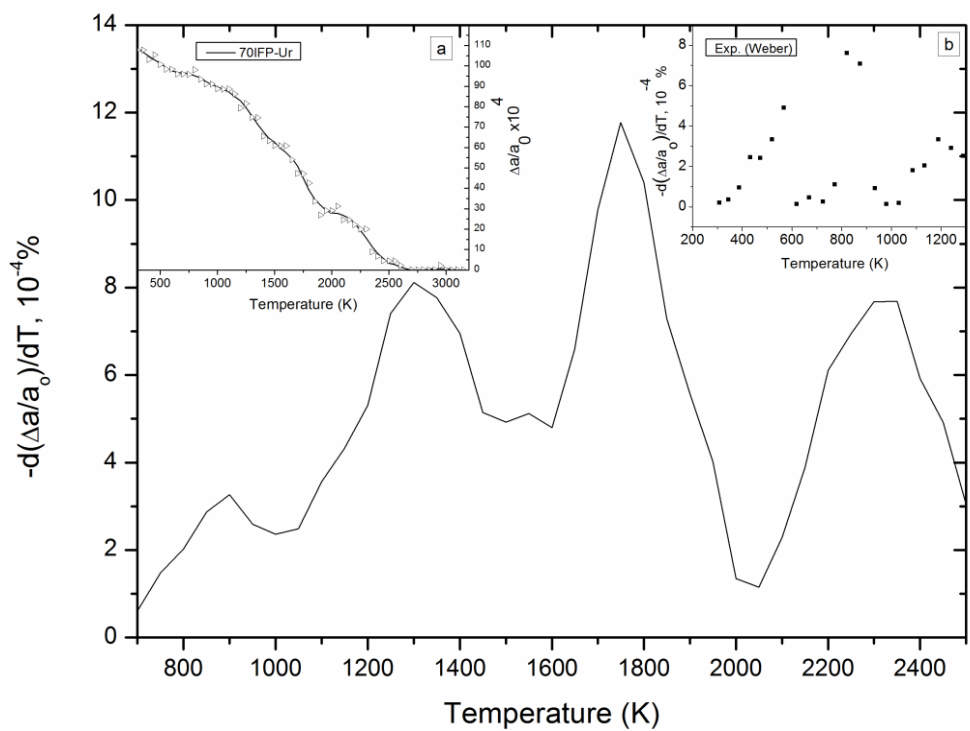

Figure 7. Derivative of the inset a. a) Smoothed data of the relative lattice change of 70 uranium IFP with temperature b) Experiment

One last comment about the Figure 7 is that there is a pre-peak prior to the "stage 1" peak at 900K. In Figure 7b, similar peak can be observed in the data of the Weber about 450 - 500K but did not commented on this.

**3.2 Isothermal annealing**

Isothermal annealing is applied to the samples with 70 IFP: for stage 1, 1300K and 1400K also for stage 2, 1750K and 1800K. Each of the simulations is performed for 10000ps (10ns). All of them is started from the previous, 50 K lower, run of 100 ps. Data is smoothed by adjacent averaging method to easily keep track and clearly demonstrate. All the results show that isothermal recovery data decreases exponentially. At low temperatures exponential decay rate is small and it increases with the temperature (Figure 8) analogous to the experimental results (inset Figure 8).

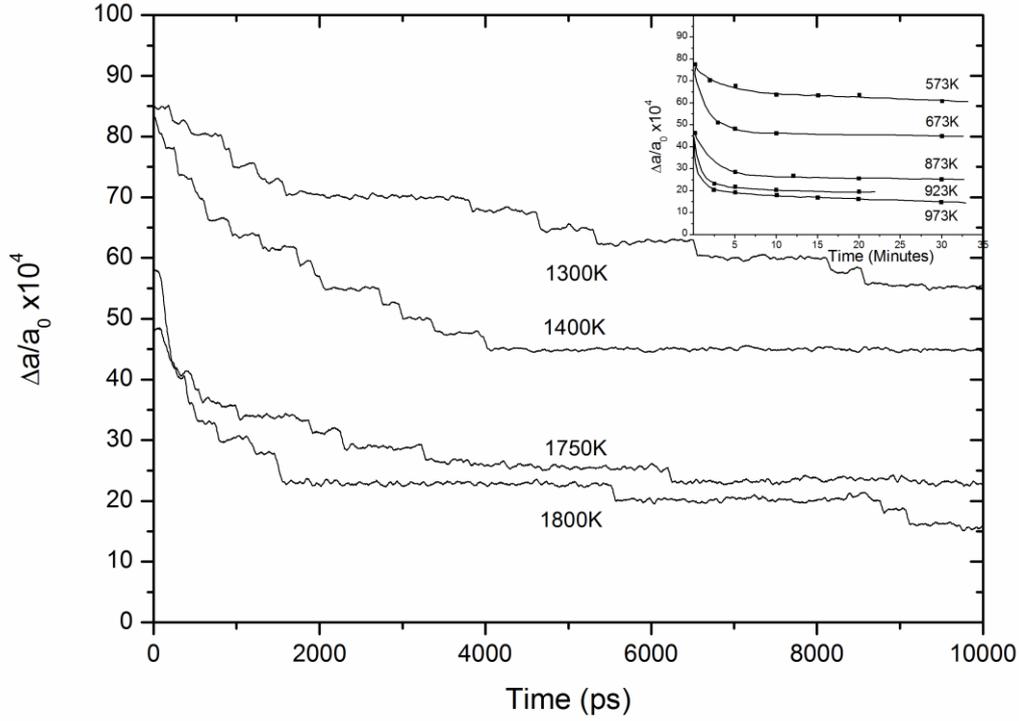

Figure 8. Isothermal annealing applied to supercell 70 uranium IFP at temperatures of the stage 1 and stage 2. Inset is the experimental results of Weber [6].

Cross-cut method [19] is applied to the data simulation data in Figure 8, briefly, parallel lines to the x-axis are drawn at random levels, t and T values are obtained, ln t versus 1/T is plotted and equation 3 is calculated,

$$\ln(t_1/t_2) = E/k(T_1^{-1} - T_2^{-1}) \tag{3}$$

This procedure is applied to stage 1 and stage 2 and activation energies are calculated as 1.85 eV and 4.0 eV respectively. Experimental calculations [6] with this method results 1.5 eV and 2.2 eV.

Weber also determined the activation energies by using the method proposed by Primak [20 - 22]. Primak had suggested an equation

$$E_0 = kT\ln(Bt) \tag{4}$$

B is the frequency factor and Weber assumed it is constant for all stages, B ≈ $10^{10}$ s$^{-1}$, t is time (30 minutes for experiment), k is the Boltzmann constant, $E_0$ is the mean activation energy which corresponds to the temperature for maximum recovery in each stage under isochronal annealing. Recovery and annealing times are very small compared to the experiment so the frequency factor must be changed, from the paper of Primak [20] it is considered that it could be about B ≈ $10^{14}$ s$^{-1}$ for simulation time scale and it is assumed constant for all stages. It is calculated as 1.5 eV, 1.945 eV and 2.55 eV and Weber found that it is 1.5 eV, 2.2 eV and 3.1 eV for all three stages respectively with this method.

## 3.3 Interstitialcy Migration Mechanism of Uranium

MD simulation method and first-principle calculations with DFT method were performed to observe and identify the interstitialcy (indirect interstitial) mechanism of uranium of the alpha-particle irradiated crystal. Energy barriers and relative energy change during recovery is calculated from these method.

### 3.3.1 MD Simulation

Indirect interstitial migration mechanism of interstitial uranium defect (obstruction type) is observed throughout the whole MD simulation, at all temperature ranges.

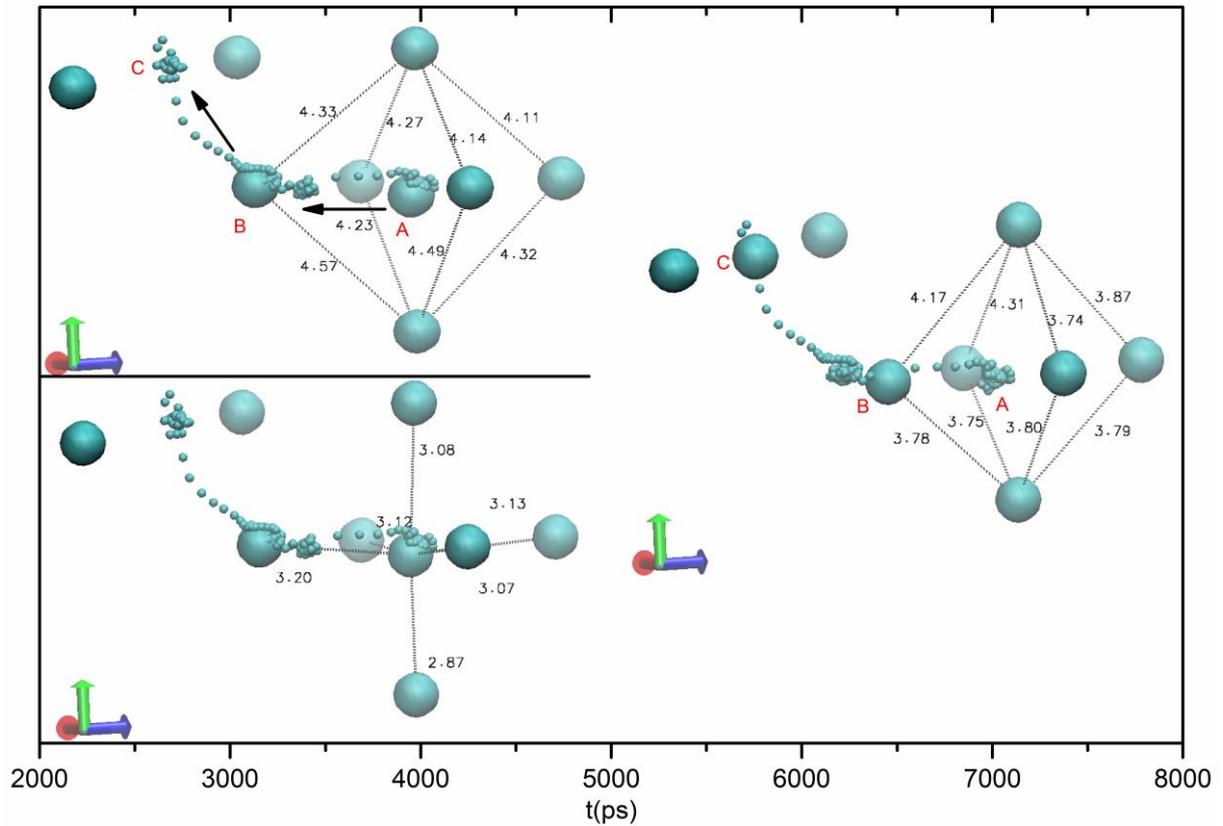

Figure 9. Instantaneous <110> view before and after the recovery of the interstitial uranium defect at the octahedral position. Uranium ions are colored with cyan. Oxygen ions are removed from the figure for clarity. At about 5000 ps recovery take place by interstitialcy mechanism.

Figure 9 is the <110> view (a little tilted around y axis) of the $UO_2$ lattice and oxygen ions are removed from the figure for clarity reasons. All the uranium ions are at their normal sites except the interstitial uranium ion at the site A which block one of the channel between uranium rows (Supplementary material). Two ions at the top left of the octahedron are displayed to indicate the position of the vacancy site C. At about 5000 ps, in Figure 10, interstitial uranium ion at site A pushes the uranium ion at site B towards the site C and move to the point B which demonstrates an interstitialcy mechanism of the recovery.

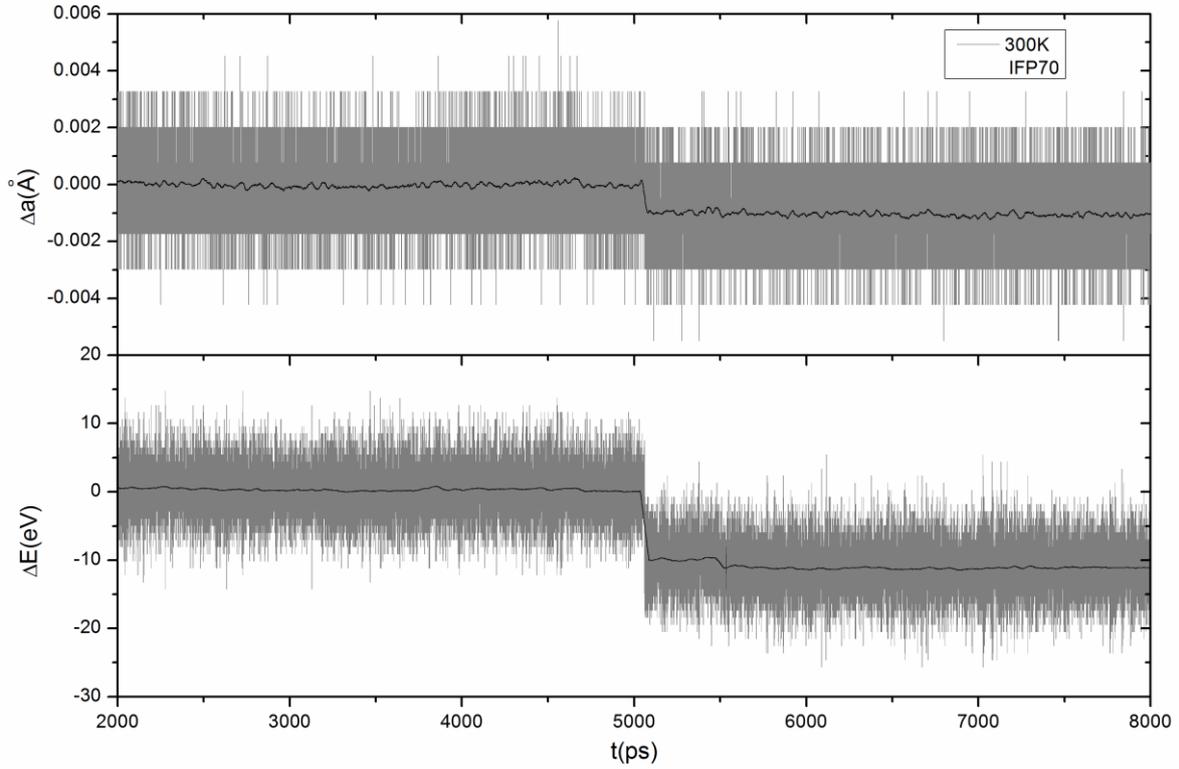

Figure 10. Change in the lattice parameter and energy with time.

Volume occupied by the octahedron decreases after the recovery where this change can be seen at the left hand side (before recovery) to the right hand side (after recovery) of the Figure 9. At the same time in Figure 10, about 5000ps, lattice parameter decreases and recovered towards the perfect crystal structure lattice parameter at 300K. Change in the lattice parameter is $\Delta a = a_{after} - a_{before} \approx -9.865 \times 10^{-4}$ Å. Similar lattice recovery steps can be seen in Figure 8 for higher temperatures. Total energy change of the crystal is $\Delta E = E_{after} - E_{before} \approx -10$ eV which means supercell is in a more favorable state after the recovery. Energy difference value corresponds for only one interstitialcy recovery of a uranium ion.

### 3.3.2 DFT Method

In this part uranium defect recovery by interstitialcy mechanism is investigated by DFT method. Interstitial (direct interstitial) migration mechanism is not considered because interstitialcy mechanism is observed in all MD simulations. Initial and final structures of the DFT is considered from the recovery mechanism of the initial and final structures of the MD simulation (similar to Figure 9). Initial structure is the interstitial uranium at octahedral site with a nearby uranium vacancy. Final structure is the perfect crystal structure. The atomic structure and the cell shape is relaxed for both initial and final states with VASP program [23-25]. Climbing-image NEB method (Cl-NEB) is applied to find the saddle point [26].

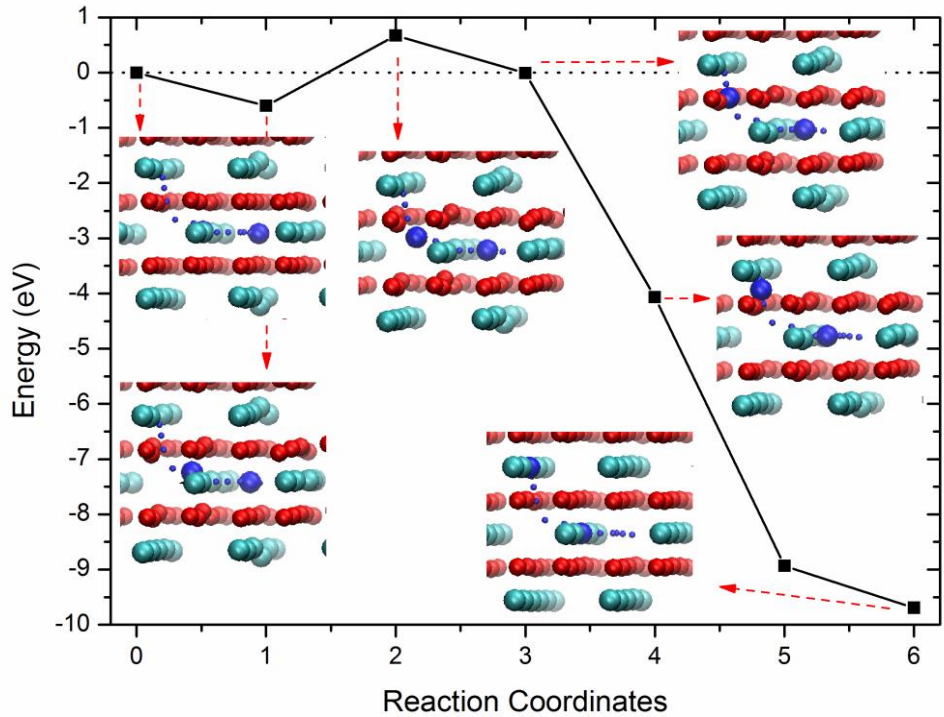

Figure 11. Relative energy of the system is calculated by Cl-NEB method as the uranium ion migrate via interstitialcy mechanism.

Migration energy value is calculated more accurately with Cl-NEB than regular NEB method because energy saddle point is found without interpolation of the energy data. Generalized solid-state NEB method [27] is also used because initial and final structures have different cell sizes and this method allows change in the lattice vectors within these two end points. Migration energy barrier value is calculated as 0.67eV. Energy difference before and after the recovery process is $\Delta E = E_{after} - E_{before} = -9.69$ eV which is very close to the MD simulation result, -10eV.

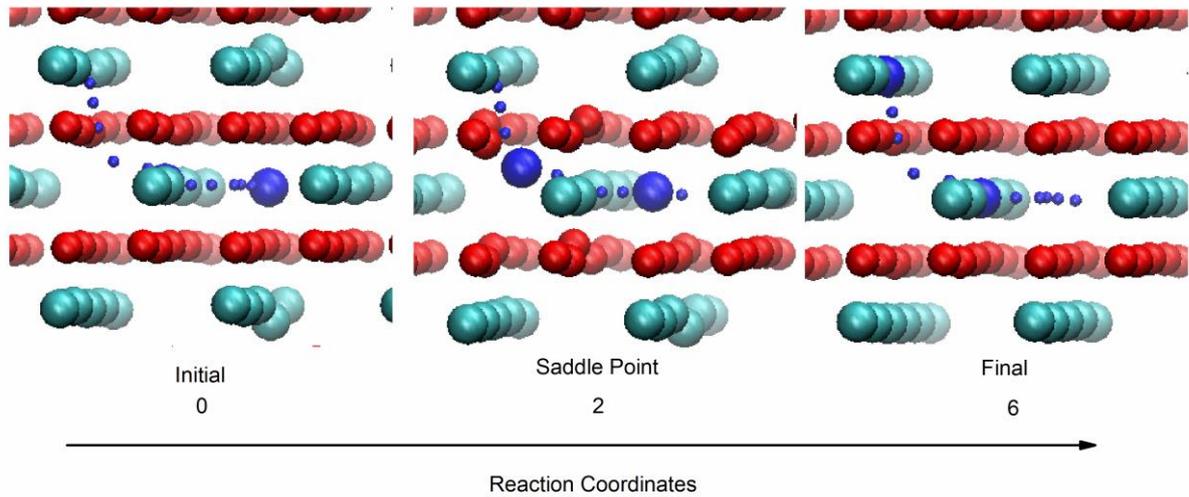

Figure 12. Initial, saddle and final <110> view of the UO$_2$ supercell calculated using the DFT method.

### 3.4 Interstitialcy Migration Mechanism of Oxygen

MD simulation method were performed to observe and identify the interstitialcy mechanism of oxygen of the neutron irradiated UO$_2$ crystal. Energy and lattice parameter change during recovery is calculated with this method. Indirect interstitial migration mechanism is observed throughout the whole MD simulations, at all temperature ranges.

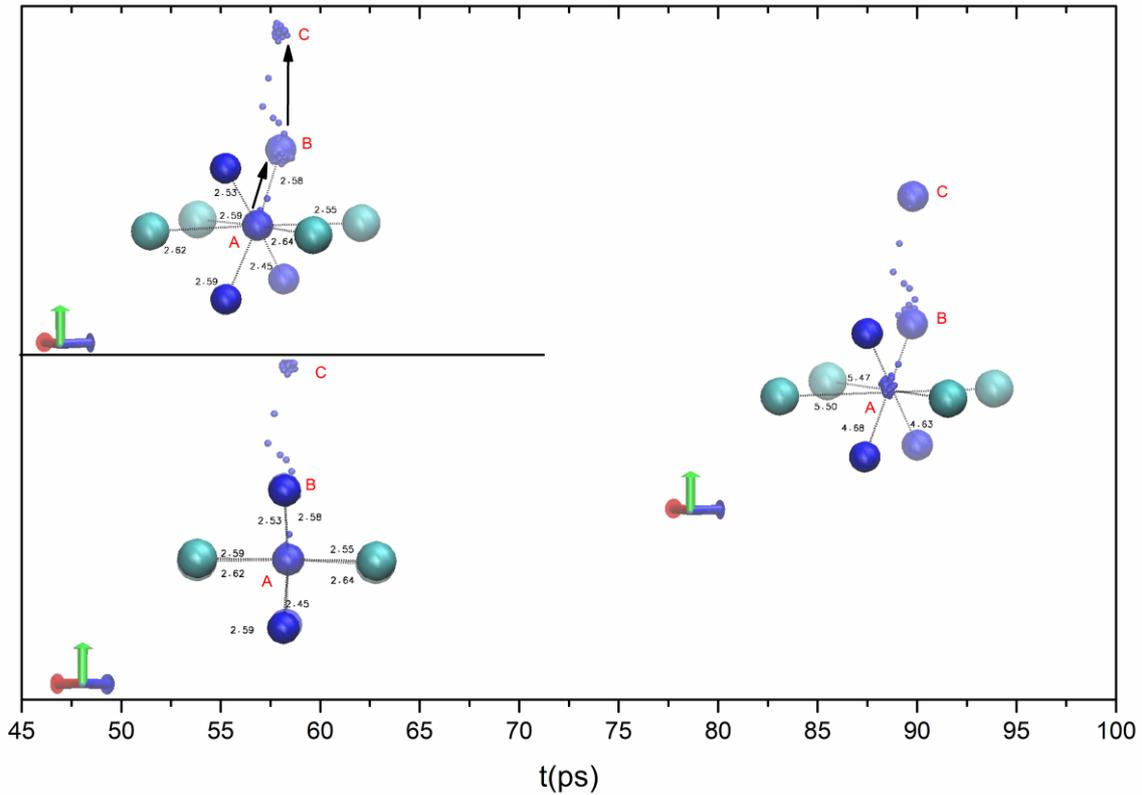

Figure 13. Instantaneous <110> view before and after the recovery of the interstitial oxygen defect. Uranium and oxygen ions are colored with cyan and blue respectively. At about 70 ps recovery take place by interstitialcy mechanism.

Figure 13 is the <110> view of the $UO_2$ lattice where the uranium ions are colored with cyan and oxygens are colored with blue. All the uranium and oxygen ions are at their normal sites except the interstitial oxygen ion at the site A on the left hand side of the figure which block the channel between uranium and oxygen rows in <110> direction. Interstitial oxygen ion is surrounded with four oxygen and four uranium ions. At about 70 ps, in Figure 13, interstitial oxygen ion at site A pushes the oxygen ion at site B towards the site C and move to the point B which demonstrates an interstitialcy mechanism of the recovery. Unlike uranium ions, positions of the normal site oxygen ions changed slightly after the recovery because oxygen ions at the interstitial position can only push outwards the other oxygen ions. Uranium ions are not affected by the interstitial oxygen ions and they remain in their positions. These could be observed by measuring the distances of the diagonal uranium and oxygen ions, as in the Figure 13.

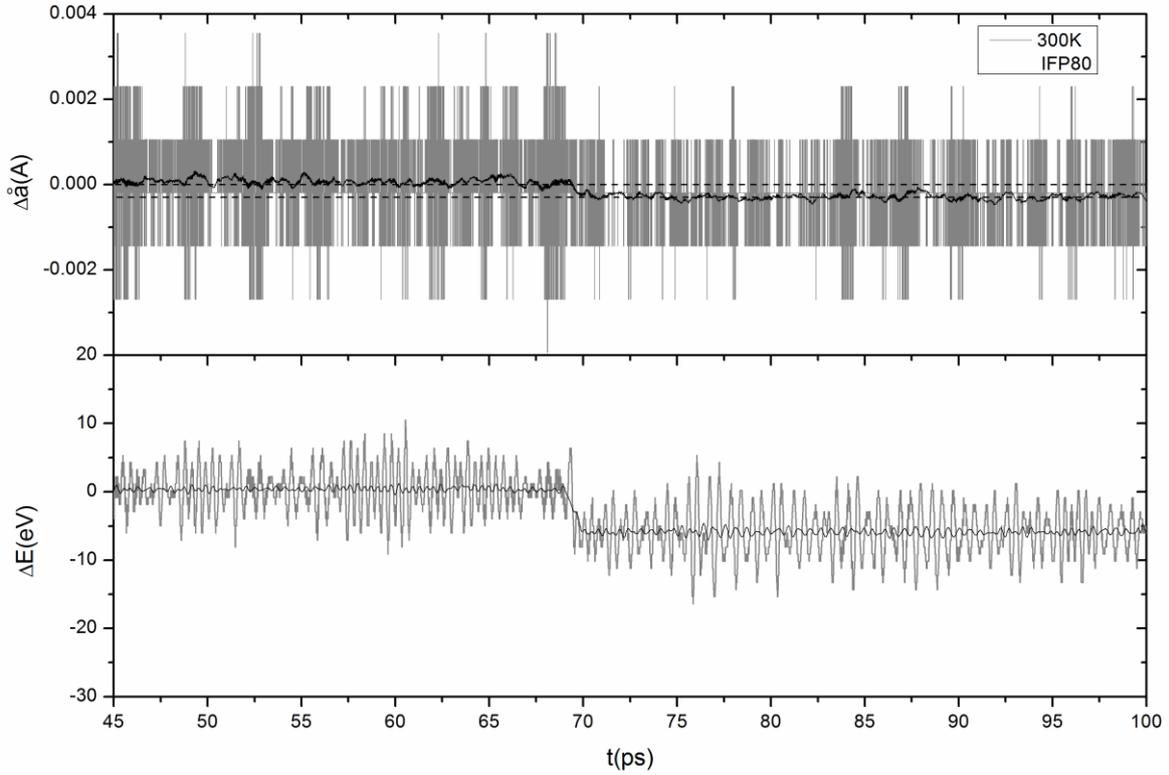

Figure 14. Change in the lattice parameter and energy with time for recovery of interstitial oxygen.

In Figure 14, both lattice and energy change is given before and after the recovery procedure of the interstitial oxygen which is demonstrated in Figure 13. At about 70ps, recovery take place in the supercell of 80 IFP oxygen at 300K. Lattice parameter change is $\Delta a = a_{after} - a_{before} \approx -3.00578 \times 10^{-4}$ Å and energy change is $\Delta E = E_{after} - E_{before} \approx -5.89$ eV.

## 4. Conclusion

In the previous study [1] swelling mechanisms of the uranium dioxide lattices was investigated. Main question were; what are the mechanisms of the swelling of lattice parameter due to radiation-induced defect ingrowth both from neutron irradiation [2] and alpha-particle irradiation [4]? Moreover why is the huge difference (almost ten folds) between the data of relative lattice expansion of these two experiments? Reliable answers had been found with the molecular dynamics simulations. Series of MD simulations had been performed with manually created oxygen and uranium Frenkel pairs in supercells of crystal uranium dioxide and then run for 100ps at 300K. Gradually increased Frenkel pairs are compared with both experimental results and it is shown that molecular dynamics data of the swelling with ingrowth of IFPs of uranium dioxide bear resemblance with the experimental swelling of crystal with irradiation dose data. In addition to that study, here in this article this phenomena is tested with a newly proposed potential, embedded atom model for uranium dioxide [13], and results are verified. Molecular dynamics simulations which are performed with both potential models proposed by Yakub et al. [12] and Cooper et al. [13] clearly shows us that reason for the swelling due to neutron irradiation is the obstruction type oxygen ions (oxygen surrounded by four oxygen and four uranium ions) and reason for the swelling due to alpha-particle irradiation is the obstruction type uranium ions (or uranium at octahedral position: uranium surrounded by six other uranium ions).

After verification of the data in the previous work, in this work, recovery of the lattice with temperature is investigated by MD simulation and again the number of these defects are the only reason for the lattice recovery. Experimental lattice recovery data both from Nakea et al. [5] and Weber [6] are compared with MD results which are in agreement. However alpha-particle irradiated $UO_2$ crystal recovery temperature points from MD simulation are over estimated about 1000K (x axis, Figure 4) but when the simulation time increased from 100ps to 500ps, simulation results are in much better agreement and over estimation value drops about 600K. This result show us that simulation time has a great impact on the recovery temperature but extending the time of the simulation does not mean that one can get better results because recovery stage points of relative lattice expansion (y axis, Figure 6) should also be taken into account and for 500ps these results are in agreement. For this reason, roughly 500 ps is a good time scale estimation for MD simulation to compare with experimental time scale results in Figure 6. Again, in Figure 8, both MD simulation and experimental results are compared and time scales between these microscopic and macroscopic systems are 10 ns and 30 minutes, respectively. Obviously difference in the time scales is from the size of the samples (size of the pellet and simulation box) so concentration of the defects inside these boundaries or relative lattice expansion should be taken into account when determining the simulation time that corresponds to experiment.

In both Nakea et al. [5] and Weber's [6] work it had been discussed which sub lattice ion is responsible for each recovery stages. Weber declared that even he found no direct information about which defect is associated with each recovery stage, he thinks the first stage is due to oxygen migration and the second stage is attributed to the uranium vacancies. Despite the huge difference in the relative lattice expansion data between alpha-particle and neutron irradiated UO$_2$ crystal data, he agrees with Nakea et al. [5] about the recovery procedure which means both alpha-particle and neutron irradiated recovery procedures of the UO$_2$ crystals are similar. There are some discussions on this but the main idea still valid for some actinide oxides up to now [7].

Our results show that neutron irradiated crystals only exhibits oxygen defects and oxygen interstitials migrate towards the oxygen vacancy points by interstitialcy mechanism. During this migration, oxygen ions repeatedly push each other towards the oxygen vacancy site. Two, three and sometimes even more oxygen ions involve in one indirect migration event. Alpha-particle irradiated crystals have both oxygen and uranium defects. As it was mentioned before obstruction type uranium ions are the reason for lattice expansion and naturally these ions are the main driving forces of the recovery procedure. However they are not dominant quantitatively which means as one uranium ion recovered by interstitialcy mechanism, two or three oxygen ions migrate to their normal sites also. This happens for the first stage of the recovery procedure. In the previous study it was shown that relative lattice change with the number of obstruction type defects are linearly dependent on each other however above a critical point, slope of linear behavior is increased because at some point oxygen ions so much distorted that a portion of them penetrate into the octahedron cage of the interstitial uranium ions. As the uranium ions are recovered, these oxygens in the octahedral cage or too much distorted (cluster) oxygen ions are also recovered (supplementary material). This critical point in the linear behavior and above correspond to the first recovery stage. It could be suggested that both oxygens and uranium defects are recovered but quantitatively oxygen ions are the dominant ones in the first stage however this recovery process starts if the interstitial uranium migrates to the vacant position even they are the minority of the defects. At this point determining the defects attributed to a recovery stage by measuring the activation energy is questionable.

One can verify these ideas by analyzing Figure 6, namely despite the 40 uranium IFP and 70 uranium IFP samples, stage 1 recovery disappeared for 20 uranium IFP because, referring the Figure 10 and Figure 11 in the previous study [1], 20 uranium IFP corresponds to the part, below the critical point of the linear behavior. In that part octahedral cages of the uranium interstitials do not possess oxygen ions. In the second and third stages only obstruction type uranium ions are recovered. In the literature recovery stages for alpha irradiated uranium dioxide crystal are oxygen interstitials, uranium vacancies and helium ions trapped in vacancies for three stages respectively which is partly similar to our results. It could be concluded that, experimental results can be interpreted as, oxygen in the first and helium in the last stage are the side effects of the recovered uranium ions from a octahedral cage to a vacancy point by interstitialcy mechanism. Finally, studies for the future could be other actinide dioxides. They also display three recovery stages [7], which give us the idea of the similar kind of defects could also be attributed to their recovery stages.